\documentstyle[11pt,newpasp,twoside,epsfig,verbatim]{article}
\index{pulsars}

\def\edcomment#1{\iffalse\marginpar{\raggedright\sl#1\/}\else\relax\fi}
\reversemarginpar

\begin{document}
\title{Electrodynamics of pulsar magnetospheres}
\author{Anatoly Spitkovsky}
\affil{KIPAC, Stanford University, P.O. Box 20450, MS 29, CA 94309}

\begin{abstract}

I review the theoretical understanding of the global structure of pulsar
magnetospheres concentrating on recent progress in force-free
electrodynamics and first-principles simulations of magnetospheres.
\end{abstract}

\section{Introduction}

It is customary to begin reviews of pulsar magnetosphere research by
calculating the ratio of the number of outstanding questions
about pulsars to the number of years passed since pulsars were 
discovered. While this ratio probably seems to decrease with time, it
does so mainly due to the inevitable growth of the denominator (now at
36), and most of the fundamental questions about pulsars are still
with us. 
This seems to scare researchers away from working on the subject. Yet,
in my view, the fact that even basic conceptual questions about
pulsars are not fully understood represents an opportunity to learn important insights
even from simplified models.  

In recent years, the subject have experienced a gradual growth of interest
which came with the reluctant realization that the structure of pulsar
magnetospheres could not be solved analytically in closed form. This and the
development of algorithms applicable to highly magnetized plasma environments
brought simulations to the forefront as the alternative tool.
I will review the current status of pulsar electrodynamics research 
concentrating on simulations, and on what they teach us about pulsars 
and the way we simulate them. 

\section{Pulsar basics} \label{pulsarbasics}

The fundamentals of pulsar energetics can be understood by considering the
spindown of a rotating conducting magnetized sphere.  Two ingredients 
are needed for its electromagnetic spindown. First is the existence of a region of rotating
magnetic field, which induces electric field $E\sim (\Omega r/c)B$,
and the second is the sweepback of magnetic field. 
This combination results in an outwards Poynting flux, with energy loss rate
$\dot{\varepsilon}={c\over 4 \pi} \int ({\bf E}\times {\bf B}) \cdot d{\bf {s}}$.
The integral can be evaluated at a fiducial radius 
$R_f$ where the sweptback toroidal field is equal to the poloidal magnetic 
field: $\dot{\varepsilon}\sim-\Omega R_f^3 B^2(R_f)$. For dipolar field being
swept back at the light cylinder $R_f\sim R_L\equiv c/\Omega$ we get
\begin{equation}
\dot{\varepsilon}\approx -{\mu^2 \Omega^4 \over c^3} f(\chi), \label{spindown}
\end{equation}
where $\mu$ is the magnetic moment, and $f(\chi)$ is a geometric
function that depends on the angle $\chi$ between the rotation and 
magnetic axes. The well-known formula for dipole spindown in vacuum
is a special case of (\ref{spindown}) with $f_{vac}(\chi)=2 \sin^2 \chi /3$. 
The argument leading to (\ref{spindown}) also holds for plasma-filled 
magnetospheres with a wind as long as the outflow is Poynting dominated. 
The physical reasons for sweepback and electric field in the 
magnetosphere are different between the vacuum and plasma-filled case. 
In vacuum the electric field is due to induction while sweepback is due to 
displacement current, with both effects disappearing for $\chi\to 0$. The 
presence of plasma, however, provides electric field due to net charge
density in the magnetosphere (e.g., corotation electric field requires
Goldreich-Julian charge density $\rho_{GJ}=-{\bf \nabla} \cdot 
({\bf \Omega}\times {\bf r} \times {\bf B})/(4 \pi c) = -{\bf \Omega} \cdot
{\bf B}/(2 \pi c)+$ relativistic correcitons). The sweepback of the magnetic
field is driven by poloidal currents, with current density 
$j_{GJ}=\rho_{GJ} c$ being of the right order of magnitude to provide 
$B_\phi=B_{poloidal}$ around the light cylinder. Unlike the vacuum case, 
the presence of plasma leads to spindown even for aligned rotators, 
$\chi=0$. How such currents are set up, circulate and close is a central 
theme of pulsar magnetospheric research. Not surprisingly, the details 
should depend on the mechanisms of plasma supply, notably on pair
creation. In the interest of keeping attention on the global picture,
I will consider only two limits: 1) plasma is supplied only from the surface 
of the star, and 2) plasma is abundant in the magnetosphere. 
The real pulsar with gaps, pair-formation fronts, etc., lies somewhere 
in between the two extremes. 

\section{Plasma supply from the surface}
Rotation of conductors in magnetic fields separates charges inside the
conductor due to Lorentz force and leads to potential differences on
the surface (unipolar induction).  Electric fields outside the
conductor develop accelerating components
($E_{||}=\bf{E}\cdot\bf{B}/|\bf{B}|$) which can extract charges from
the surface if the work function is sufficiently small. This
extraction lead Goldreich and Julian (1969) to propose a
charge-separated model of magnetosphere filled with plasma at particle
density equal to $\rho_{GJ}/e$.  The viability of this model has been
questioned over the years (e.g., Holloway 1973), and an unsolved issue
of principle is whether an aligned rotator with particle emission only from
the surface would spin down (and hence form an active current
system).

The evolution of a rotating sphere with negligible work function has
been simulated by several groups (Krauss-Polstorff and Michel 1984,
1985; Smith, Michel \& Thacker, 2001;
 Petri, Heyvaerts \& Bonazolla 2002a; Spitkovsky \& Arons 2002, 
and in preparation, hereafter SA). The first three groups used variations of an
iterative procedure that constructed axisymmetric 2D charge-separaed 
magnetospheres by releasing charges from the surface, finding their 
equilibrium locations
and then iterating locations of other charges until convergence was found.
In our work (SA) we use a more general approach by utilizing particle-in-cell
(PIC) method for electromagnetic plasma simulation. We represent plasma 
as a collection of mactoparticles and solve inhomogeneous
Maxwell equations with currents provided by motion of macroparticles while
computing their motion in self-consistent fields. Special care is taken to 
represent rotating conducting sphere as a boundary condition. The method 
is 3D and is geared towards simulating dynamic situations 
and not limited to static equilibria. 

So, what happens when a conducting sphere (or star) with a dipole 
magnetic field is rotated? 
Unipolar induction generates electric field corresponding to a central 
monopolar charge plus quadrupolar surface charge.
This field pulls negative charges over the poles (${\bf{\Omega}} || {\mathbf{\mu}}$)
and positive charges in the equatorial region (fig. 1a-c) and has natural trapping regions 
where $\bf{E}\cdot\bf{B}=0$. Once particles start to populate the magnetosphere they 
provide quadrupolar space charge which tends to cancel the accelerating electric 
field at the surface, reducing the surface charge and particle injection, approaching
a quasi-steady state. This state, reproduced by simulations 
from all groups, consists of negative ``domes'' of charge at magnetic poles, and 
a positive torus around the equator, with electric gap in between. Inside the 
plasma regions the accelerating electric field is zero, and the plasma density 
is on the order of, but not exactly, $\rho_{GJ}$. The total extent of the 
charge-separated configuration is several radii of the star, and the magnetosphere 
is dead: there is no longitudinal current, no magnetic sweepback, and no spindown. 

\vskip -.3in
\begin{figure}[hbt]
\unitlength = 0.0011\textwidth
\begin{center}
\hspace{1\unitlength}
\begin{picture}(140,200)(0,15)
\put(0,0){\makebox(126,200){ \epsfxsize=138\unitlength \epsfysize=220\unitlength
\epsffile{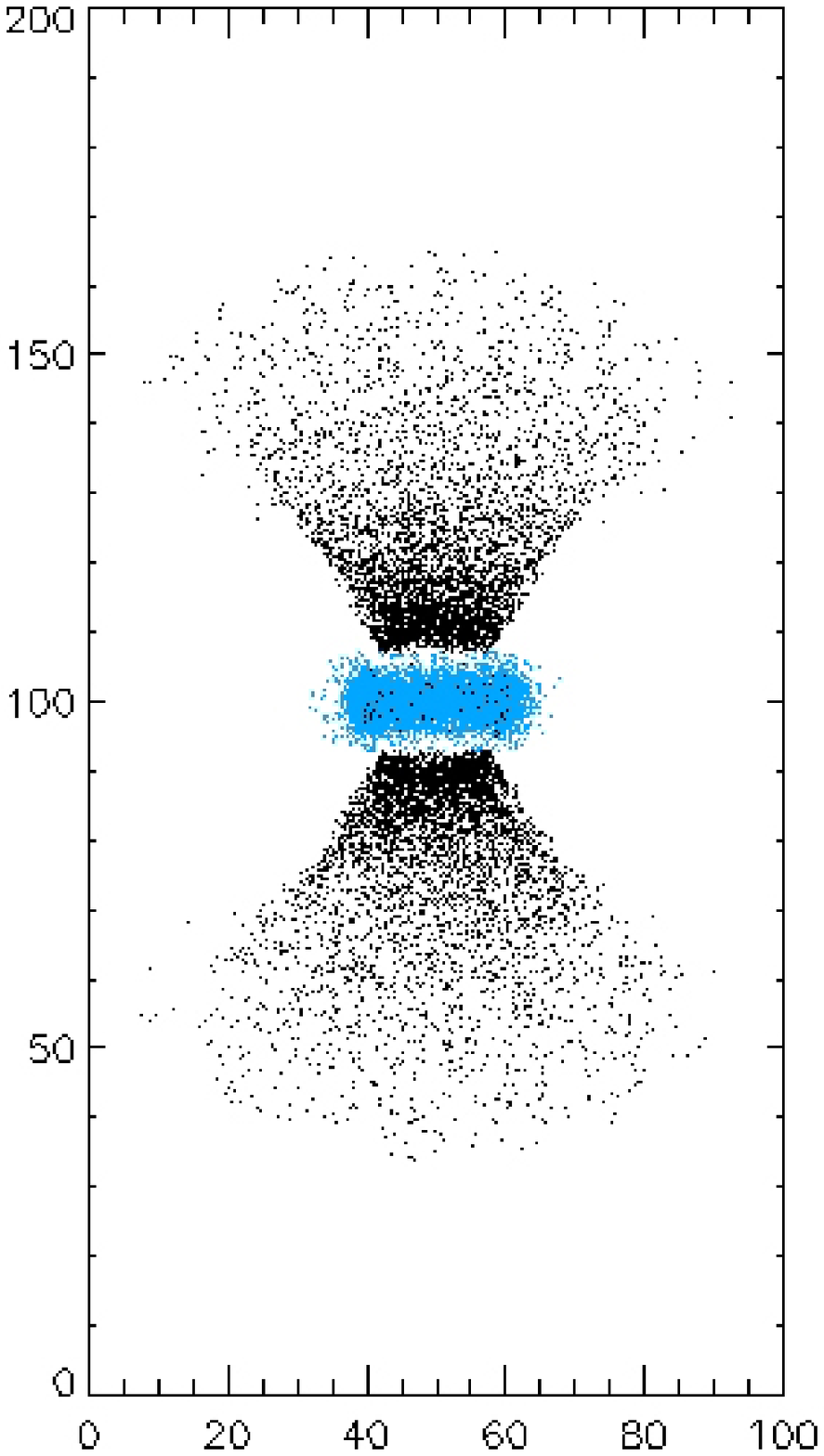}}}
\put(10,170){\makebox(0,0){\tiny (a)}}
\end{picture}
\hspace{1\unitlength}
\begin{picture}(140,200)(0,15)
\put(0,0){\makebox(126,200){\epsfxsize=138\unitlength \epsfysize=220\unitlength
\epsffile{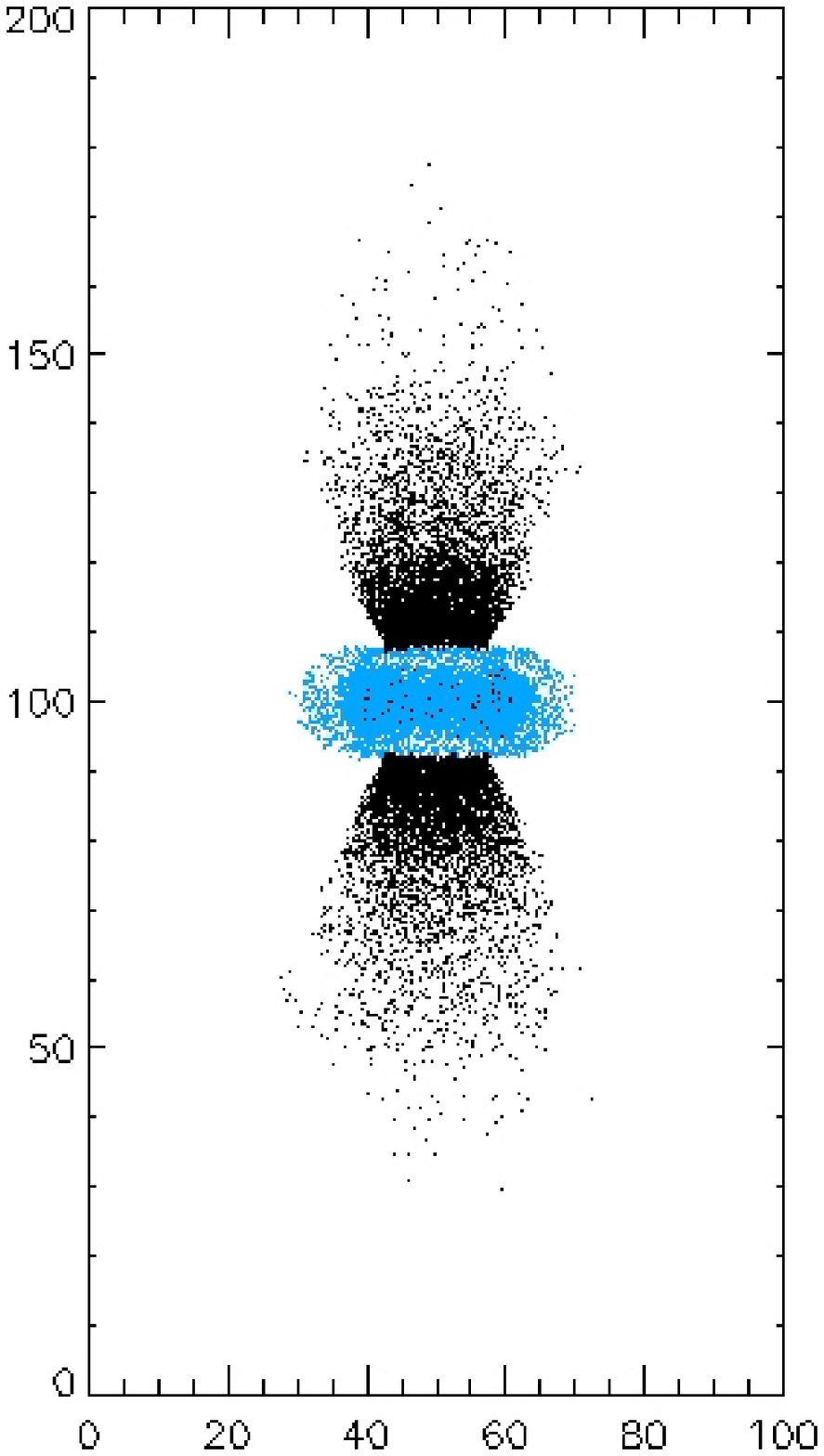}}}
\put(10,170){\makebox(0,0){\tiny (b)}}
\end{picture}
\hspace{1\unitlength}
\begin{picture}(140,200)(0,15)
\put(0,0){\makebox(126,200){\epsfxsize=138\unitlength \epsfysize=220\unitlength
\epsffile{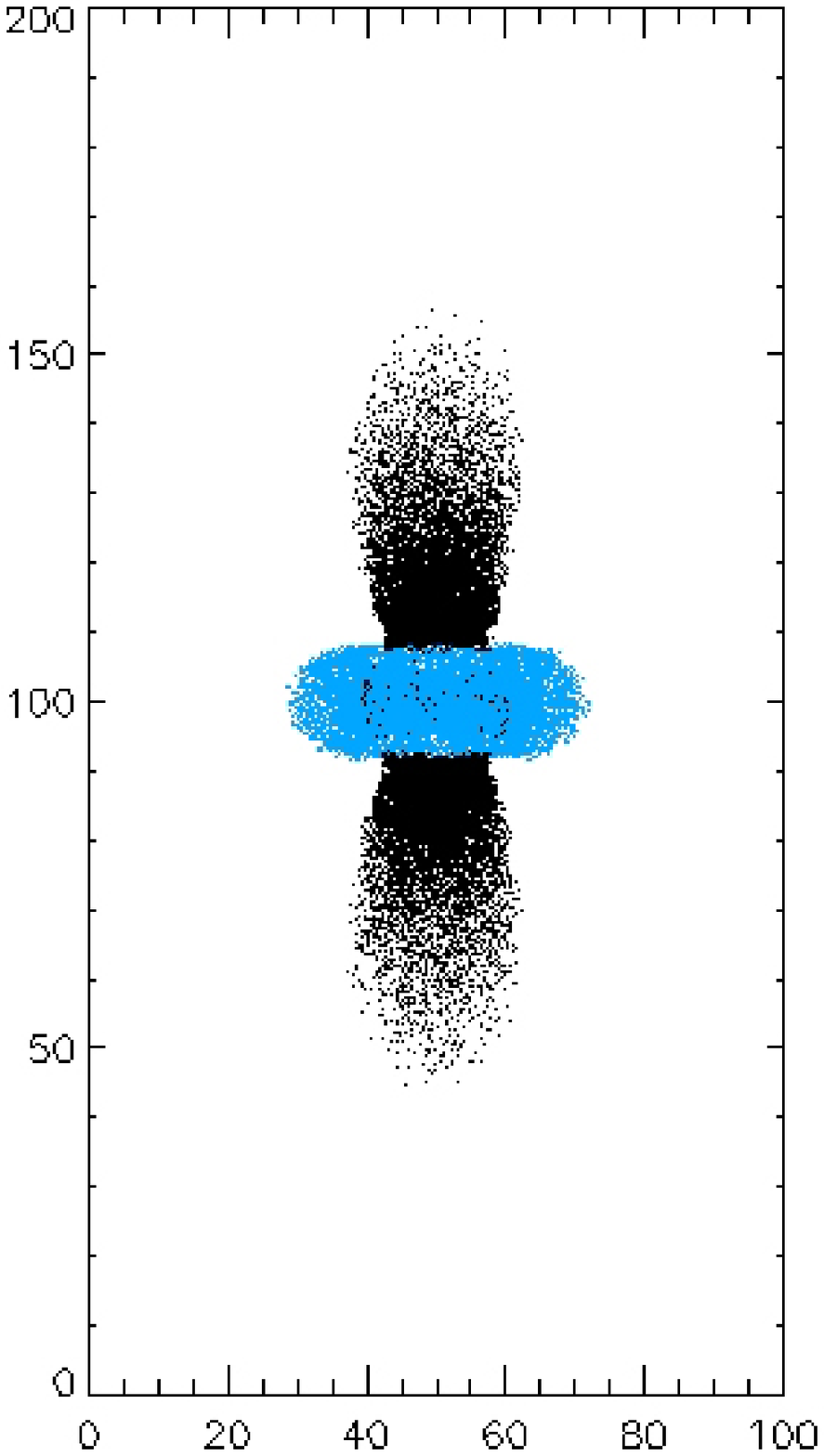}}}
\put(10,170){\makebox(0,0){\tiny (c)}}
\end{picture}
\hspace{1\unitlength}
\begin{picture}(200,200)(0,15)
\put(0,0){\makebox(250,200){\epsfxsize=275\unitlength \epsfysize=220\unitlength
\epsffile{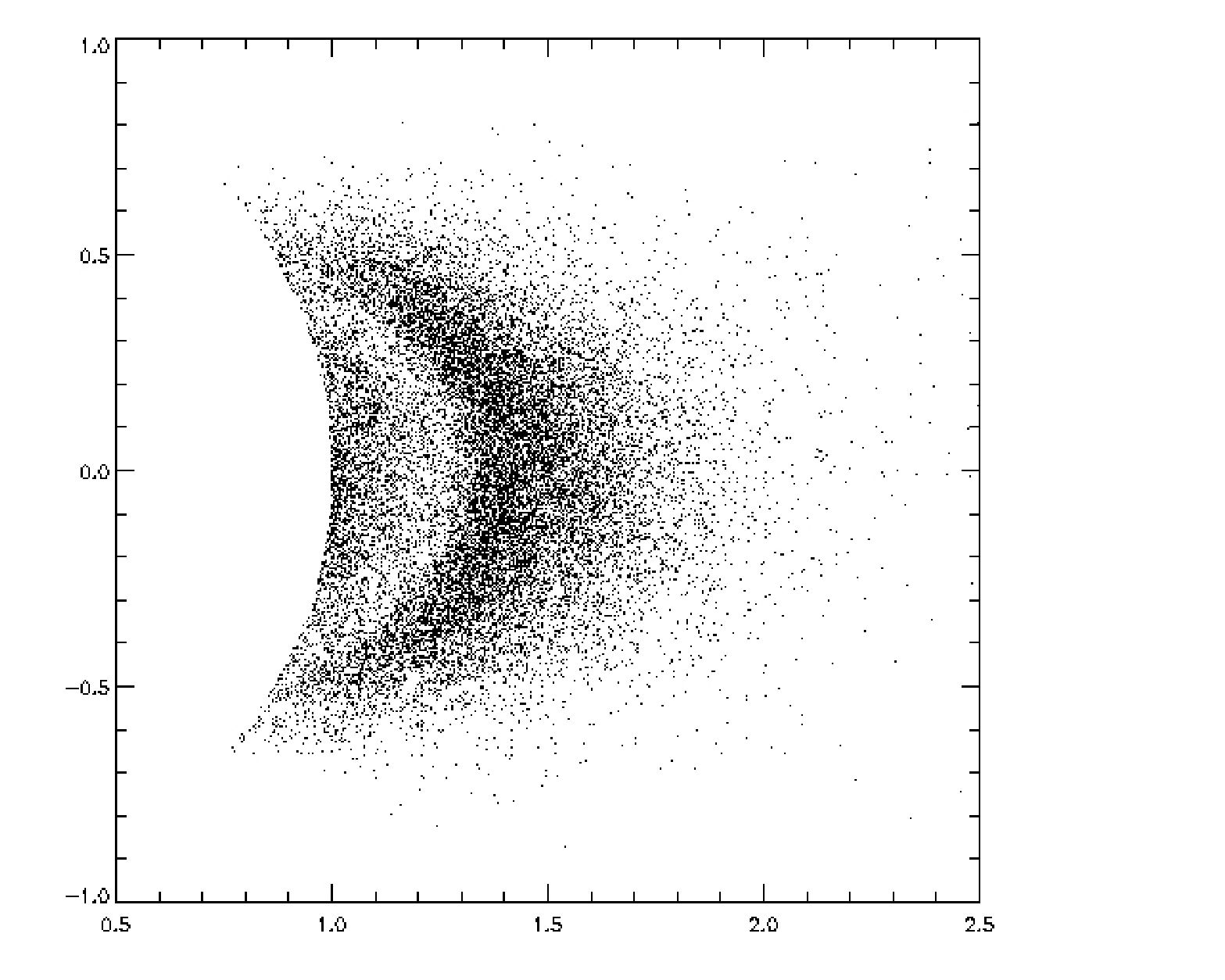}}}
\put(0,170){\makebox(0,0){\tiny (d)}}
\end{picture}
\hspace{1\unitlength}
\begin{picture}(200,200)(0,15)
\put(0,0){\makebox(200,200){\epsfxsize=220\unitlength \epsfysize=220\unitlength
\epsffile{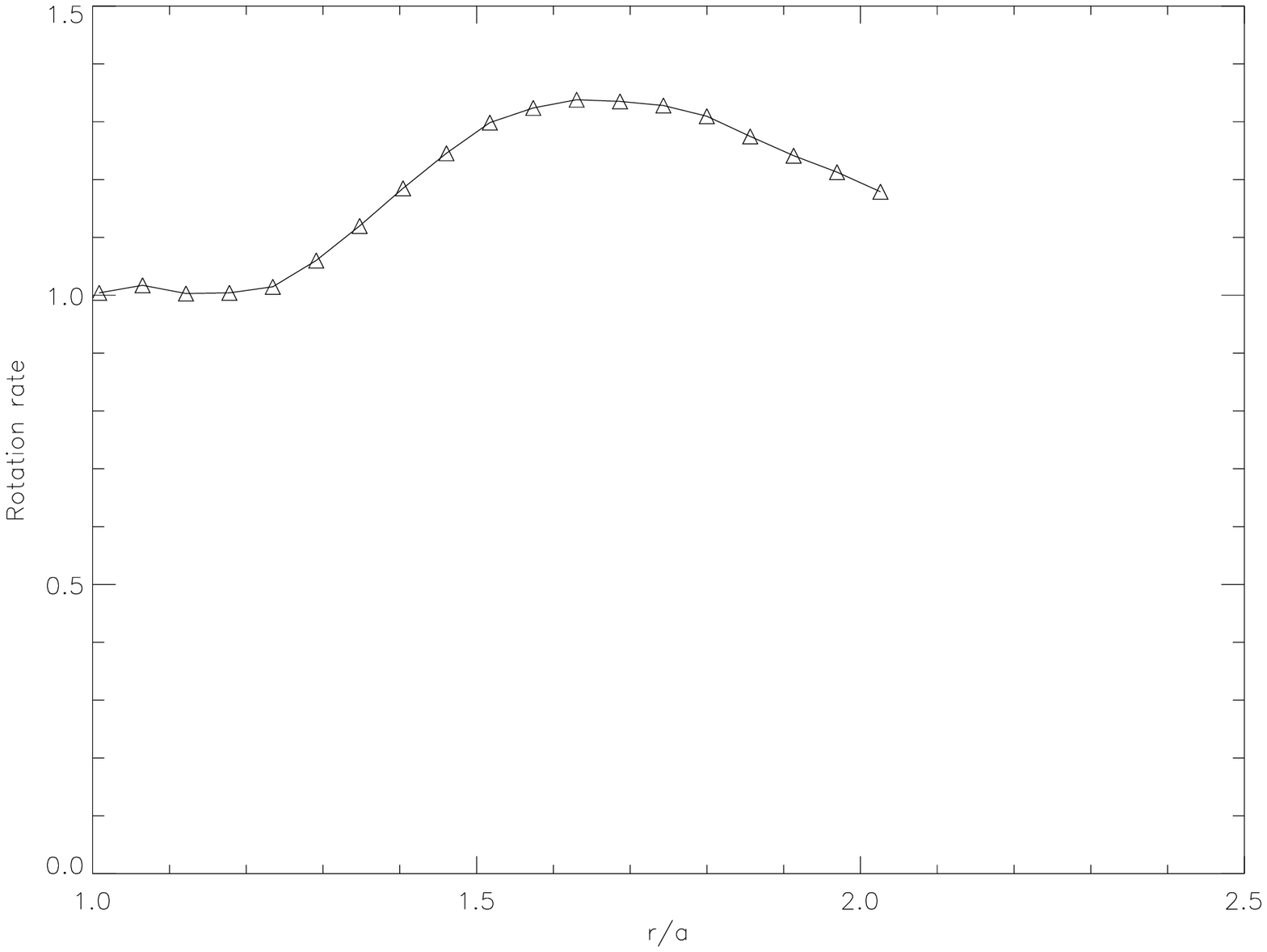}}}
\put(0,170){\makebox(0,0){\tiny (e)}}
\end{picture}
\end{center}
\vskip -.15in
\caption{
Formation of charge-separated magnetosphere with particle emission from the surface of a conducting sphere with dipole  field. }
\end{figure}

At least that is the story for axisymmetric simulations. In 3D the quasi-static
configuration is actually unstable.
Some fieldlines that go through the equatorial torus also pass through regions of
vacuum gaps. These fieldlines are not equipotentials and hence plasma on them is not
corotating with the star. Figures 1d-e show the cross-section through the positive
torus and a plot of the plasma rotation rate (in units of stellar $\Omega$) along 
an equatorial radius. Differential 
rotation in non-neutral plasma leads to shear instability known as the 
diocotron instability. 
In it the azimuthal charge density perturbations grow by transporting charges 
accross the field lines. The transport is due to the $\bf{E}\times\bf{B}$ drift 
in perturbed azimuthal electric field and poloidal magnetic field, and was 
observed in simulations of SA. The growth rate for the instability is on the order
of several stellar periods comparable to linear instability analysis of  
Petri et al (2002b).
The instability feeds off the differential rotation of the plasma and tries 
to establish corotation by moving charge density. The nonlinear simulations (fig.
2) show the equatorial torus growing with time and approaching GJ density. 

\begin{figure}[hbt]
\unitlength = 0.0011\textwidth
\begin{center}
\hspace{1\unitlength}
\begin{picture}(200,200)(0,15)
\put(0,0){\makebox(200,200){ \epsfxsize=200\unitlength \epsfysize=200\unitlength
\epsffile{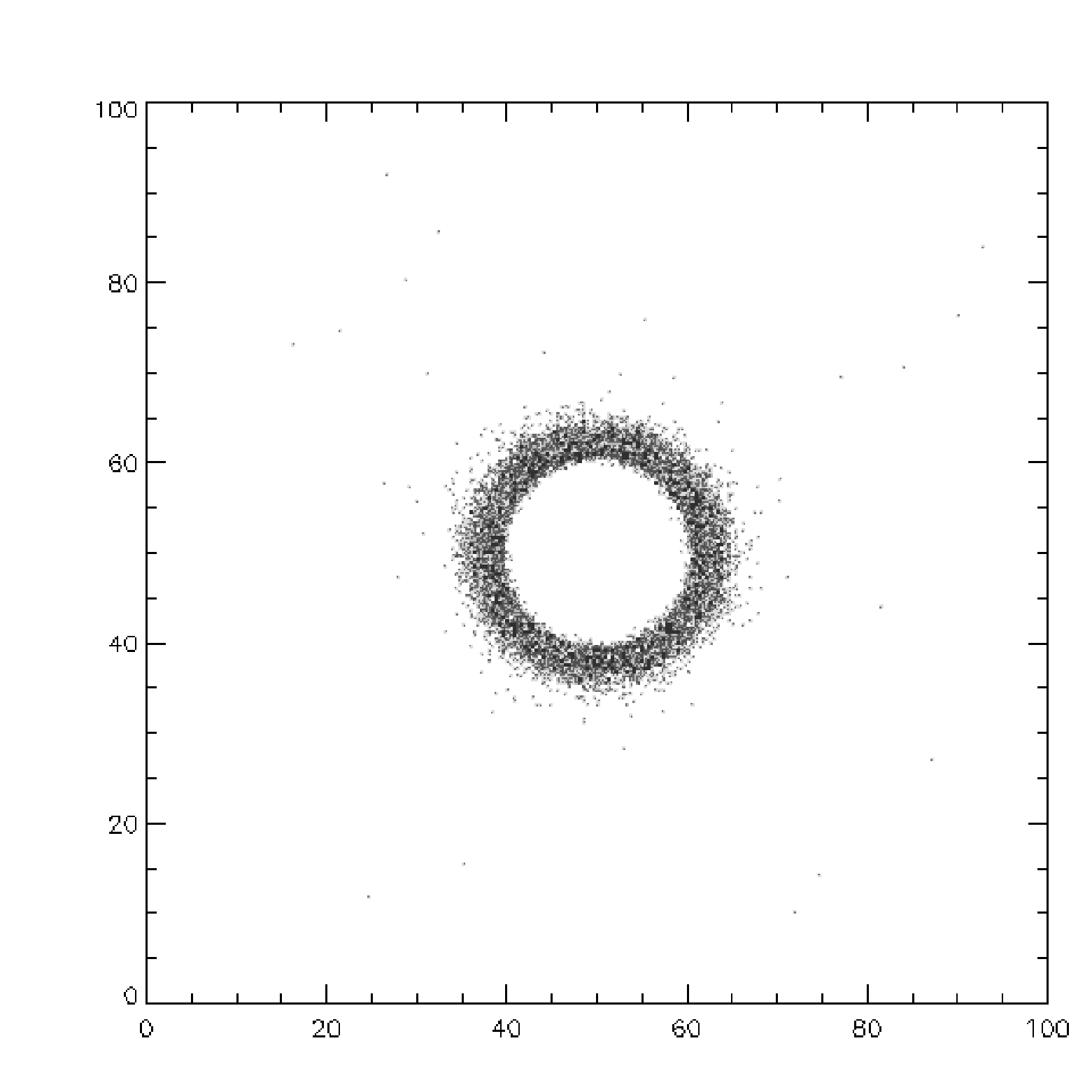}}}
\end{picture}
\hspace{10\unitlength}
\begin{picture}(200,200)(0,15)
\put(0,0){\makebox(200,200){\epsfxsize=200\unitlength \epsfysize=200\unitlength
\epsffile{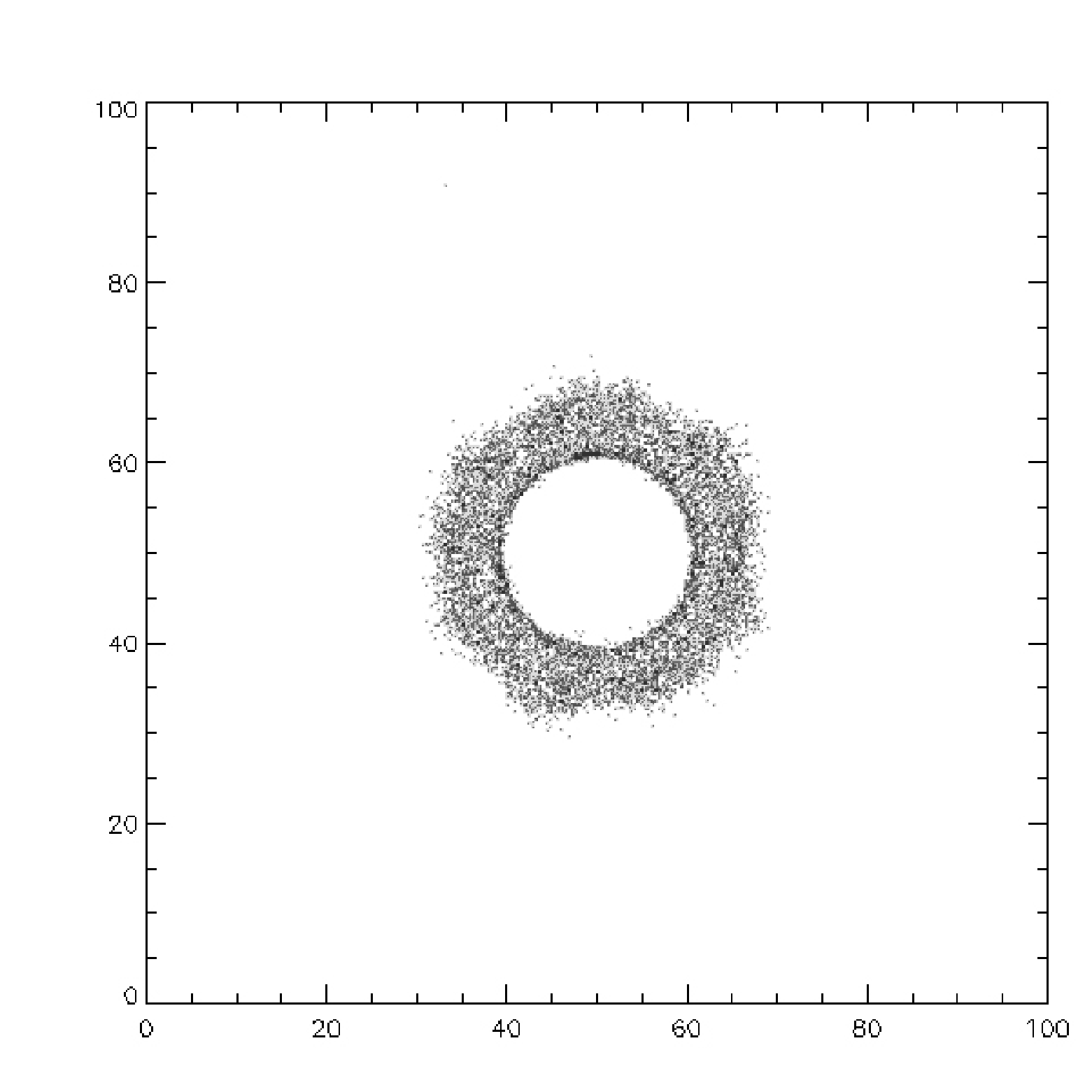}}}
\end{picture}
\hspace{10\unitlength}
\begin{picture}(200,200)(0,15)
\put(0,0){\makebox(200,200){\epsfxsize=200\unitlength \epsfysize=200\unitlength
\epsffile{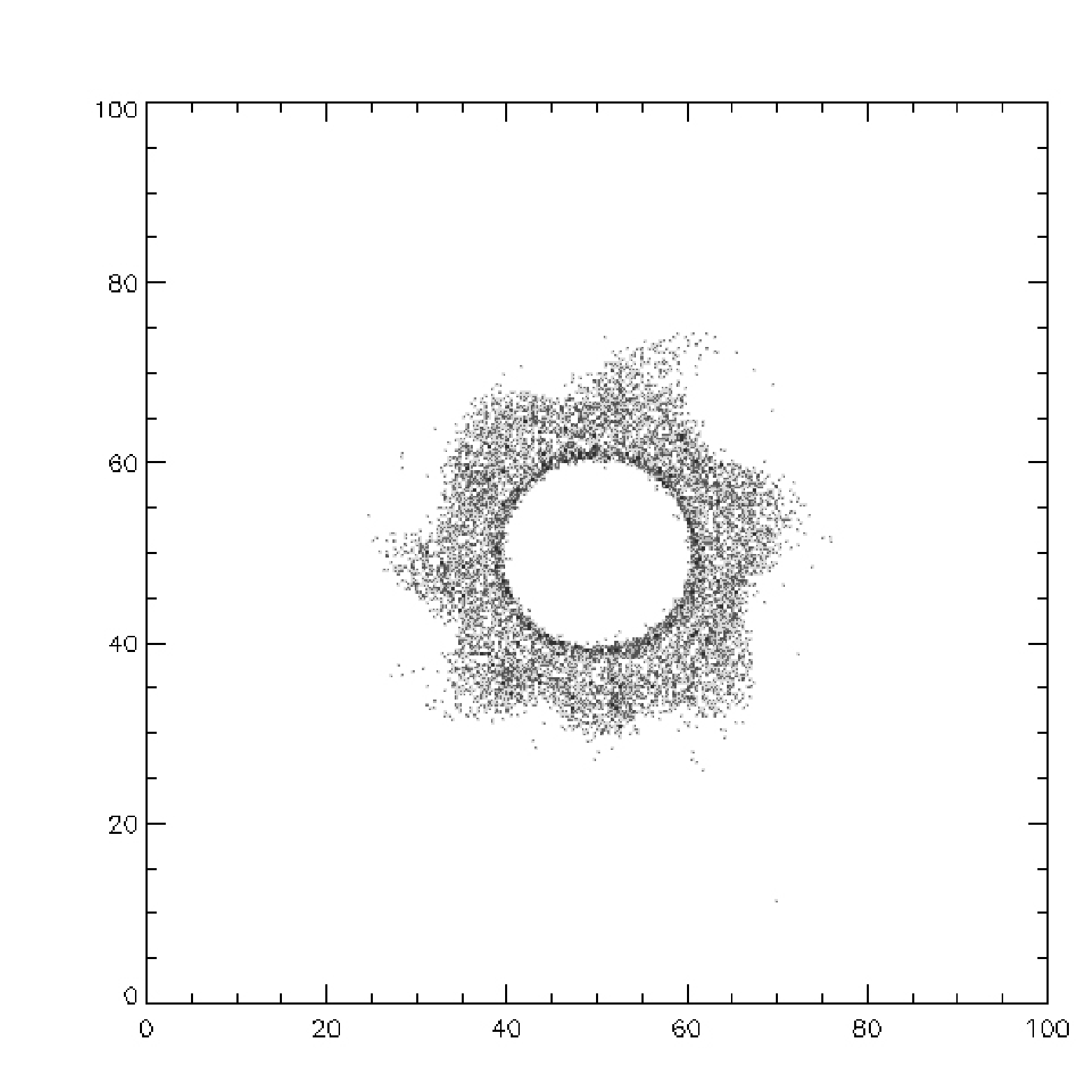}}}
\end{picture}
\hspace{10\unitlength}
\begin{picture}(200,200)(0,15)
\put(0,0){\makebox(200,200){\epsfxsize=200\unitlength \epsfysize=200\unitlength
\epsffile{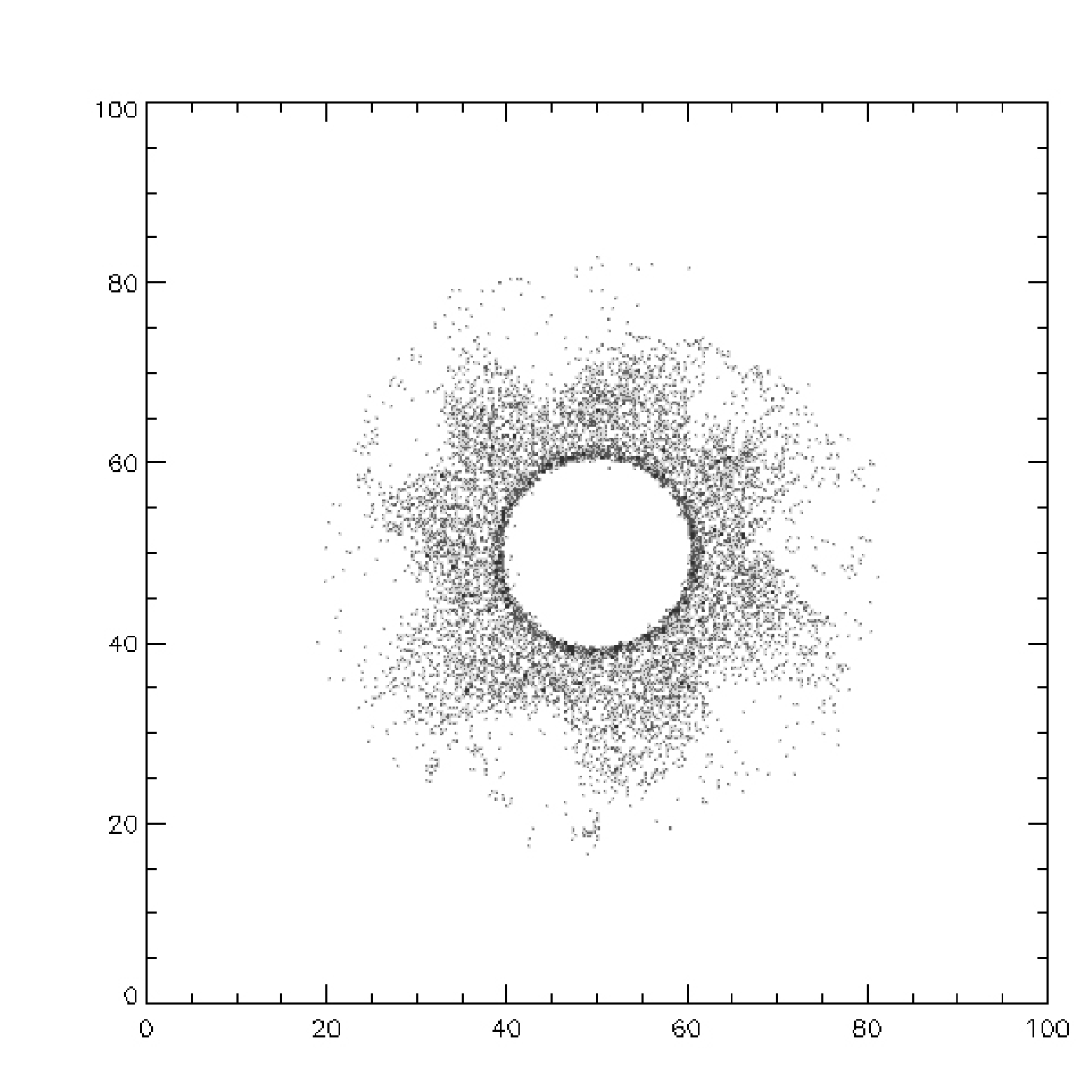}}}
\end{picture}
\end{center}
\vskip -.2in
\caption{
Development of diocotron instability in the equatorial disk of charge-separated aligned rotator.}
\end{figure}

The fact that in 3D the static ``dome-disk'' electrospheres are unstable and display
evolution to corotation suggests that dismissing charge-separated magnetospheres
as dead is a bit premature. Having a corotating closed zone is a crucial part
of the GJ model, because rotation of charges in the closed zone provides the 
toroidal current needed to modify and open the poloidal field lines.
Radial transport of particles in the equator is a current that changes the charge
on the star, and the domes expand to compensate. The present simulations have not
yet convincingly addressed the limit when the domes and disk expand out to the 
light cylinder, when the instability turns from essentially electrostatic to 
electromagnetic. 
In the absence of a definitive simulation, I will speculate on 
a possible scenario that can bring a charge-separated magnetosphere alive. 
Initial expansion of the corotating closed zone proceeds upto the light cylinder
where charge pileup causes regions with $E>B$ to form. 
The particles decouple from the magnetic field and leave generating the 
current in the 
equator. This current connecting through the star causes current flow in 
the negative domes, perhaps with particles slingshot away in the vicinity of 
the light cylinder as well. 
This initial longitudinal current flow modifies the magnetic geometry, 
causing sweepback, and gradually particles both in the equatorial region and 
in the domes recouple to the fieldlines, now with a wind outflow.
The diocotron instability and the development
of transient accelerating regions is speculatively used here to jump-start the 
current flow. 
\vskip -.1in
\begin{figure}[hbt]
\unitlength = 0.0011\textwidth
\begin{center}
\hspace{1\unitlength}
\begin{picture}(200,200)(0,15)
\put(0,0){\makebox(200,200){ \epsfxsize=176\unitlength \epsfysize=300\unitlength
\epsffile{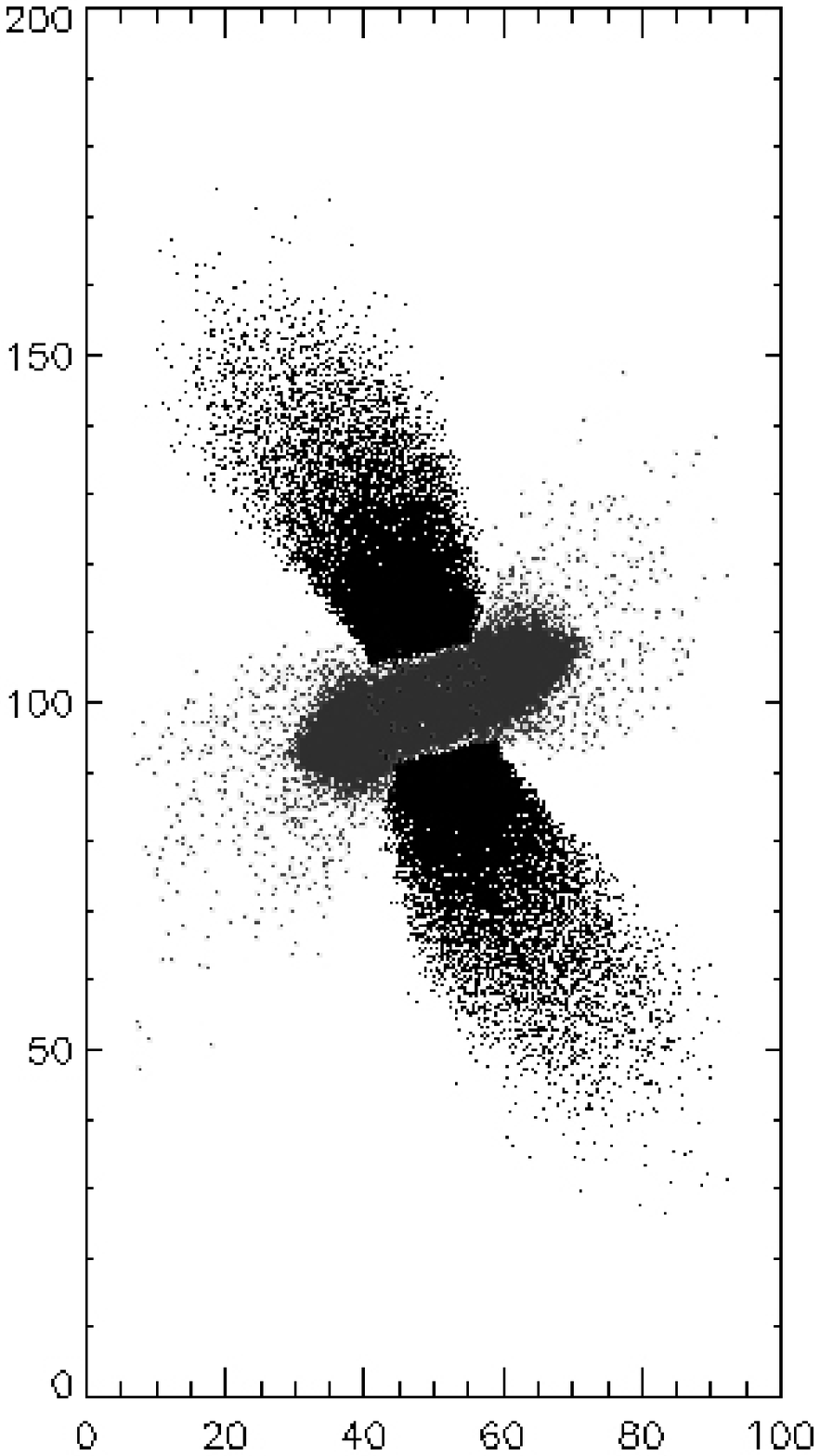}}}
\end{picture}
\hspace{1\unitlength}
\begin{picture}(200,200)(0,15)
\put(0,0){\makebox(200,200){\epsfxsize=176\unitlength \epsfysize=300\unitlength
\epsffile{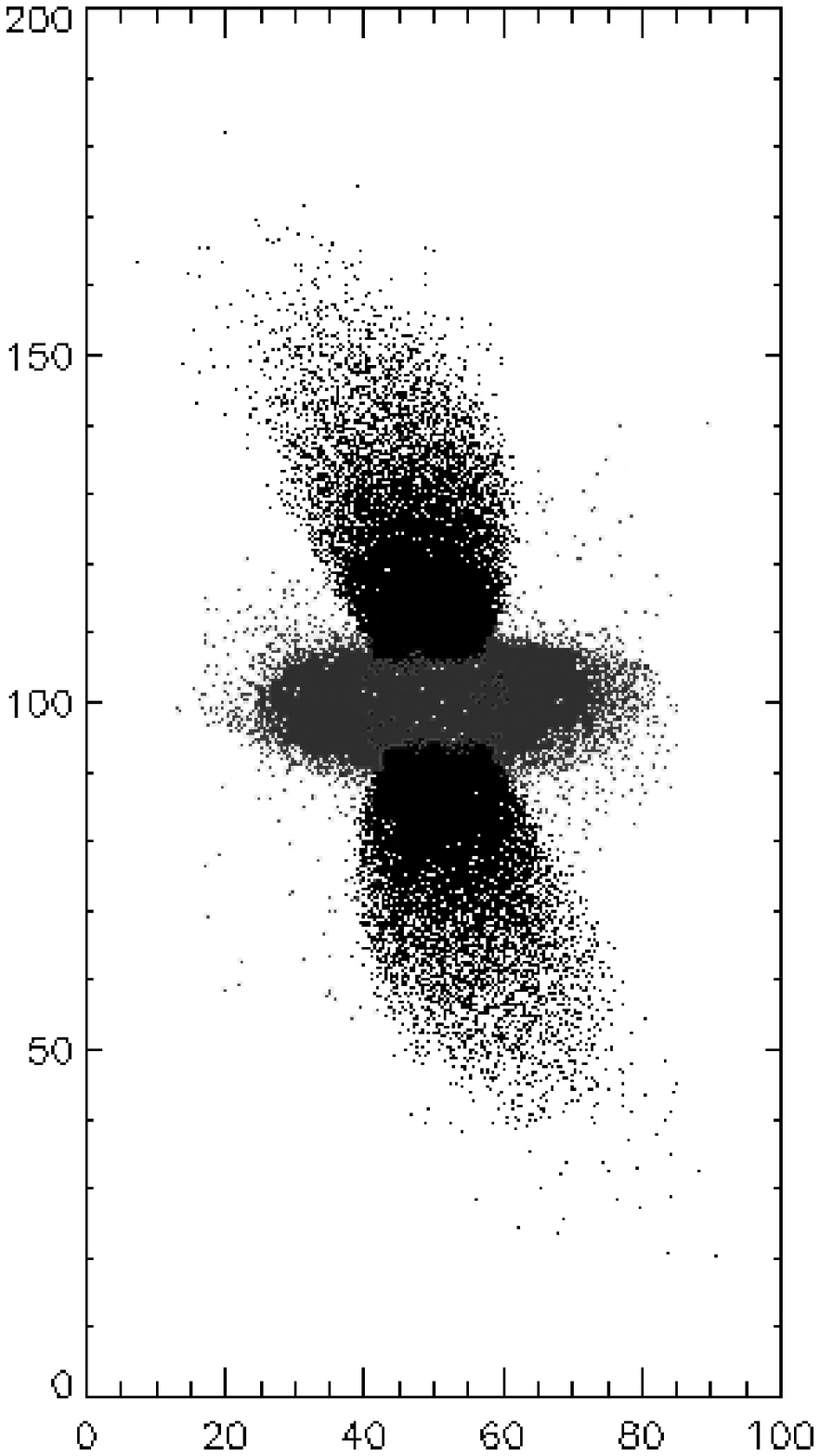}}}
\end{picture}
\hspace{1\unitlength}
\begin{picture}(200,200)(0,15)
\put(0,0){\makebox(200,200){\epsfxsize=176\unitlength \epsfysize=300\unitlength
\epsffile{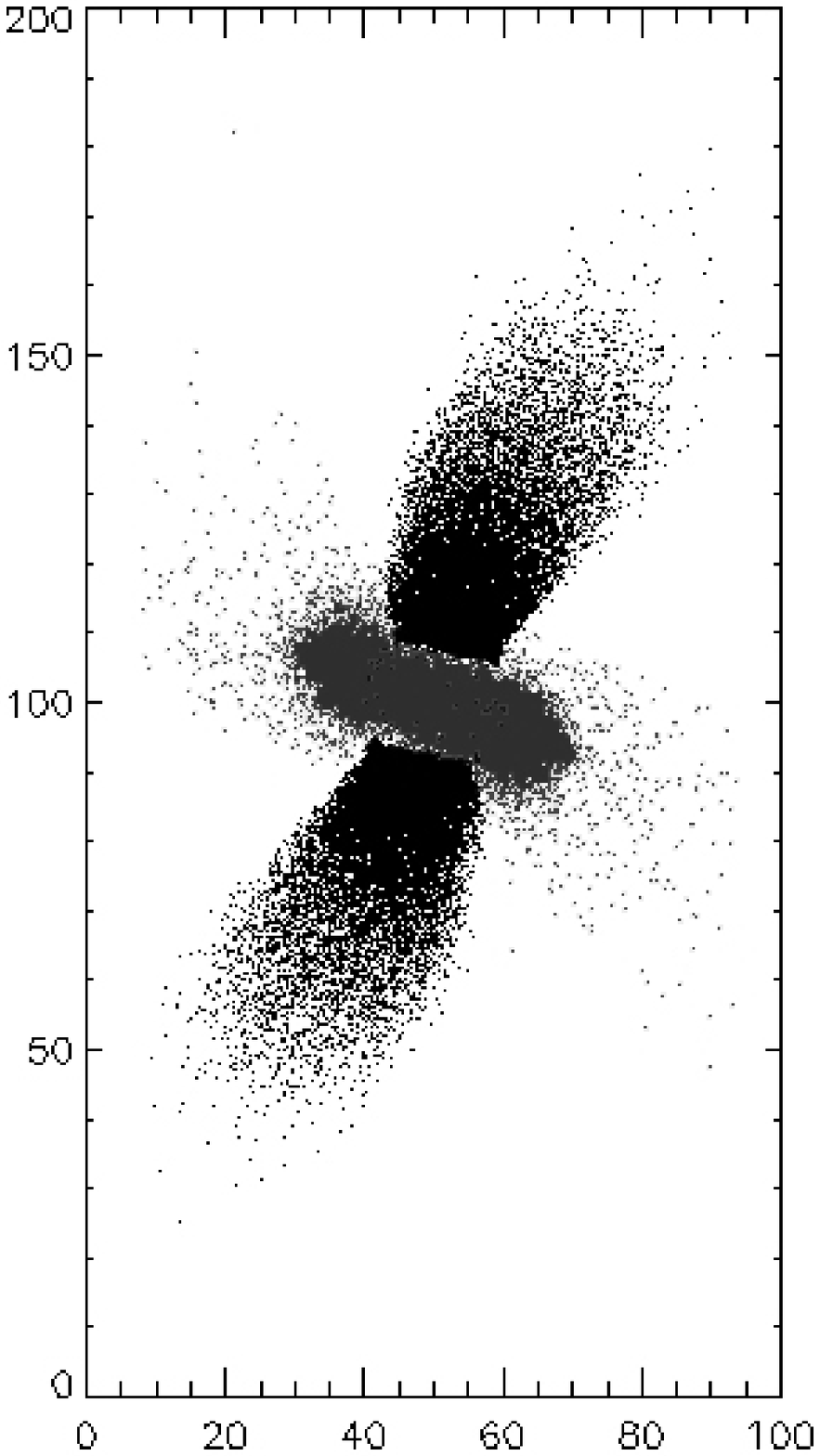}}}
\end{picture}
\hspace{1\unitlength}
\begin{picture}(200,200)(0,15)
\put(0,0){\makebox(200,200){\epsfxsize=176\unitlength \epsfysize=300\unitlength
\epsffile{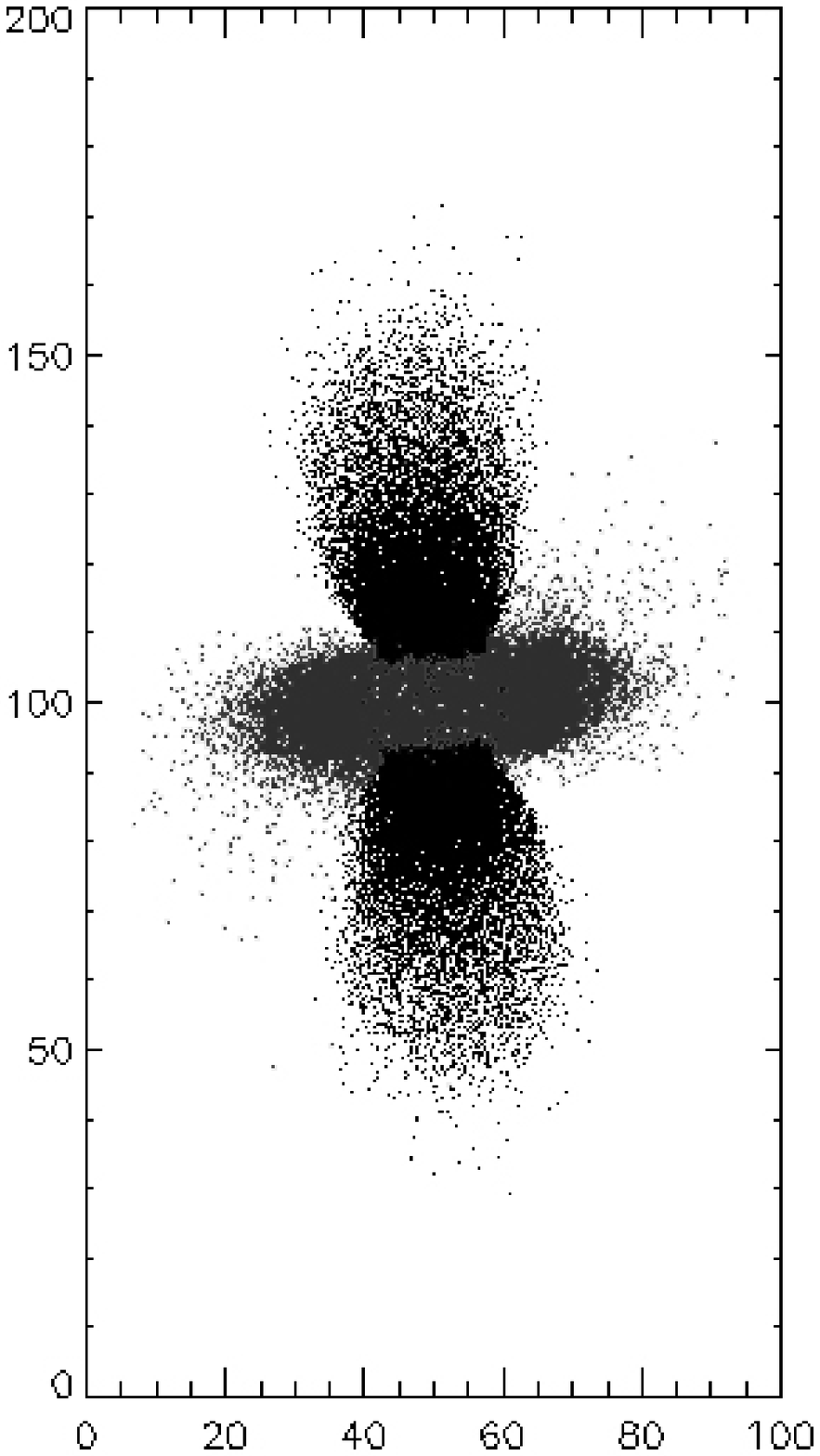}}}
\end{picture}
\end{center}
\caption{
Development of electrosphere and diocotron instability for $30^\circ$ oblique 
rotator: 3D view.}
\end{figure}


Another possibility for jump-starting a pulsar, which is still virtually 
unexplored, is the effect of magnetic inclination.
Our preliminary simulations (figure 3) show that even for oblique ($30
^\circ$) rotator the ``dome-disk'' electrosphere is formed, albeit with torus 
distorted into disjoint halves. The diocotron instability still takes place,
and the torus grows. Peculiar to the oblique rotator is the wave pressure from the
dipole spindown, which increases towards $R_{LC}$. The wave pressure 
initially drives a small fraction of particles away, potentially starting a current flow. 
The study of long-term evolution of oblique rotators is currently underway. 
Even if charge-separated magnetospheres turn out to be dead in the end, the charge
transport due to diocotron instability may still operate in the closed
zone of an active pulsar. The field lines that connect to the star through
a null surface will lose corotation if their charge content is perturbed. The 
diocotron instability may then transport charge across the fieldlines to return
the closed zone to corotation.

\section{Force-free magnetospheres}

It is commonly thought that the problems with charge-separated
magnetospheres would be readily solved by introducing sufficient amount of 
quasineutral plasma, most likely due to pair creation. This plasma
is able to quickly short out accelerating fields so that ideal
MHD condition $\bf{E}\cdot\bf{B}=0$ holds everywhere. The structure of 
the magnetosphere is then determined by the equilibrium of forces acting on 
the plasma. In the limit of small plasma inertia and temperature, the right
hand side of the momentum equation becomes
$\rho {\bf{E}}+{1\over c}{\bf{j}}\times {\bf{B}}=0$,
or ``relativistic force-free''. Significant effort was directed over the
years to finding a steady-state axisymmetric solution for the structure of 
magnetosphere of an aligned rotator subject to these constraints. Introducing
flux function $\Psi$ such that poloidal field is ${\bf B_p}=\nabla \Psi \times \hat{\phi}/r_c$ and writing corotation electric field as 
${\bf{E}}={r_c \Omega \over c} {\bf{B}_p} \times \hat{\phi}$, the force-free constraint 
can be reduced to
\vskip -.1in
\begin{equation}
(1-x^2) [{\partial^2 \Psi \over \partial x^2}+{\partial^2 \Psi \over \partial z^2}]-
{1+x^2 \over x } {\partial \Psi \over \partial x}=-I(\Psi) I'(\Psi).
\label{pulseq}
\end{equation}
\vskip -.05in
Here, $x\equiv r_c/R_L$ and $z\equiv z_c/R_L$ are scaled cylindrical 
coordinates, and $I(\Psi)$ is an unknown function proportional to the 
polodial current enclosed by the flux surface $\Psi$. Known as the ``pulsar
equation'' (Michel 1973a) (\ref{pulseq}) is a nonlinear second order 
elliptic equation with a regular singularity at the light cylinder $x=1$. 
The current distribution $I(\Psi)$ is a priori unknown and should be determined
as part of the self-consistent solution. 
Analytic solutions for pulsar equation are known only for a monopolar magnetic
field (Michel 1973a), and for a corotating dipole magnetosphere without 
a current (Michel 1973b). A great advance has recently been made by 
Contopoulos, Kazanas \& Fendt (1999, hereafter CKF) who solved the pulsar 
equation numerically for dipole magnetic field. They used an elliptic 
solver to solve
eq. (\ref{pulseq}) both inside and outside the light cylinder for a trial
current distribution. In general, such a solution has an unphysical kink 
in the magnetic field at the light cylinder. A clever 
iterative algorithm that adjusted the current function to minimize the 
discontinuity allowed CKF to find a solution that
passed smoothly through the light cylinder. This solution, shown in Fig. 4a,
has a corotating closed zone and an open wind zone, with poloidal fieldlines 
asymptotically becoming monopolar, and toroidal field dominating at infinity. 
The last open field line corresponds to 
$\Psi_{open}=1.36 \Psi_{pc}$ where $\Psi_{pc}\equiv \mu/R_L$ is the flux 
through the polar cap for unperturbed dipole field. 

The magnetosphere supports an active current system: current is flowing 
along the open field lines to infinity, and the bulk of return current flows
in the equatorial current sheet and on the boundary of closed and open zones, 
although a $5\%$ fraction is distributed 
along a set of open field lines near equator 
($1.08 \Psi_{pc} < \Psi < 1.36 \Psi_{pc}$).
 The CKF solution is intuitively
pleasing, and should be analyzed further for stability and used as a global 
model for studying plasma processes in pulsars. However, the questions of
validity and uniqueness of CKF solution still linger. At the time of this 
writing, the CKF result could not be numerically reproduced by several groups.
At issue is the representation of a delta-function return current at the 
closed-open field line interface. CKF reported smoothing this transition with
a Gaussian, however, repeating this prescription have not yielded convergence.
Instead, Ogura \& Kojima (2003) and the present author found that there exists
another smooth solution with $\Psi_{open}=1.66 \Psi_{pc}$ when the 
delta-function current is ignored (figure 4b). Although an unphysical solution
with $E>B$ beyond the light cylinder, it demonstrates that convergence to a 
smooth solution is not a guarantee of uniqueness and should be taken with caution. 
\vskip -.3in
\begin{figure}[hbt]
\unitlength = 0.0011\textwidth
\begin{center}
\hspace{1\unitlength}
\begin{picture}(300,300)(0,15)
\put(0,0){\makebox(300,300){ \epsfxsize=300\unitlength \epsfysize=300\unitlength
\epsffile{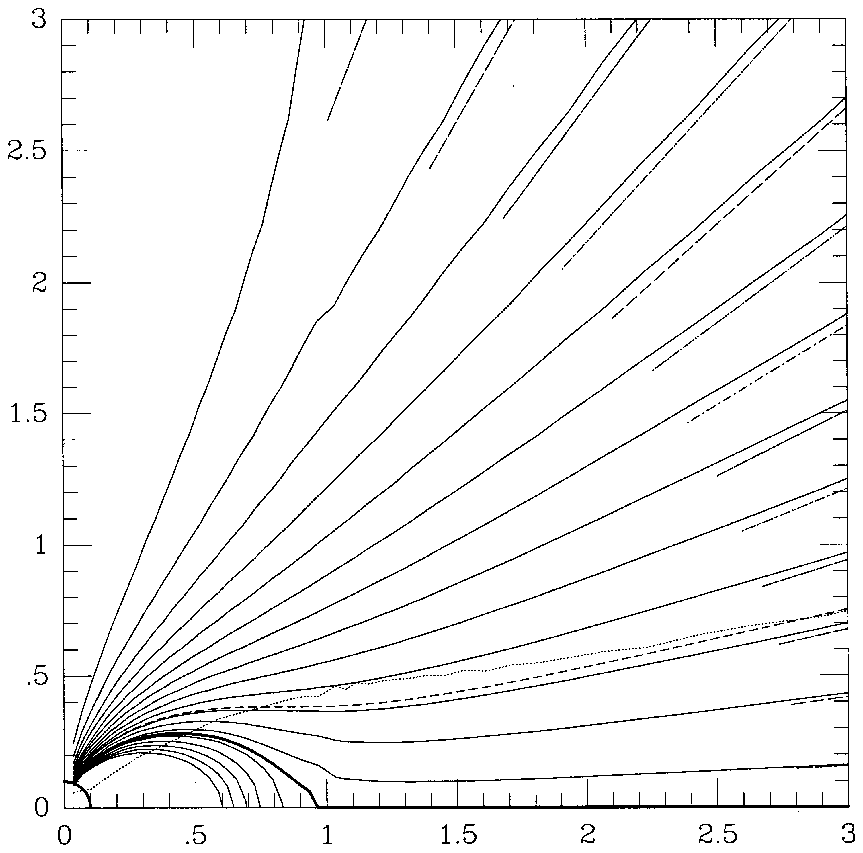}}}
\put(0,270){\makebox(0,0){\tiny (a)}}
\end{picture}
\hspace{40\unitlength}
\begin{picture}(300,300)(0,15)
\put(0,0){\makebox(300,300){\epsfxsize=300\unitlength \epsfysize=300\unitlength
\epsffile{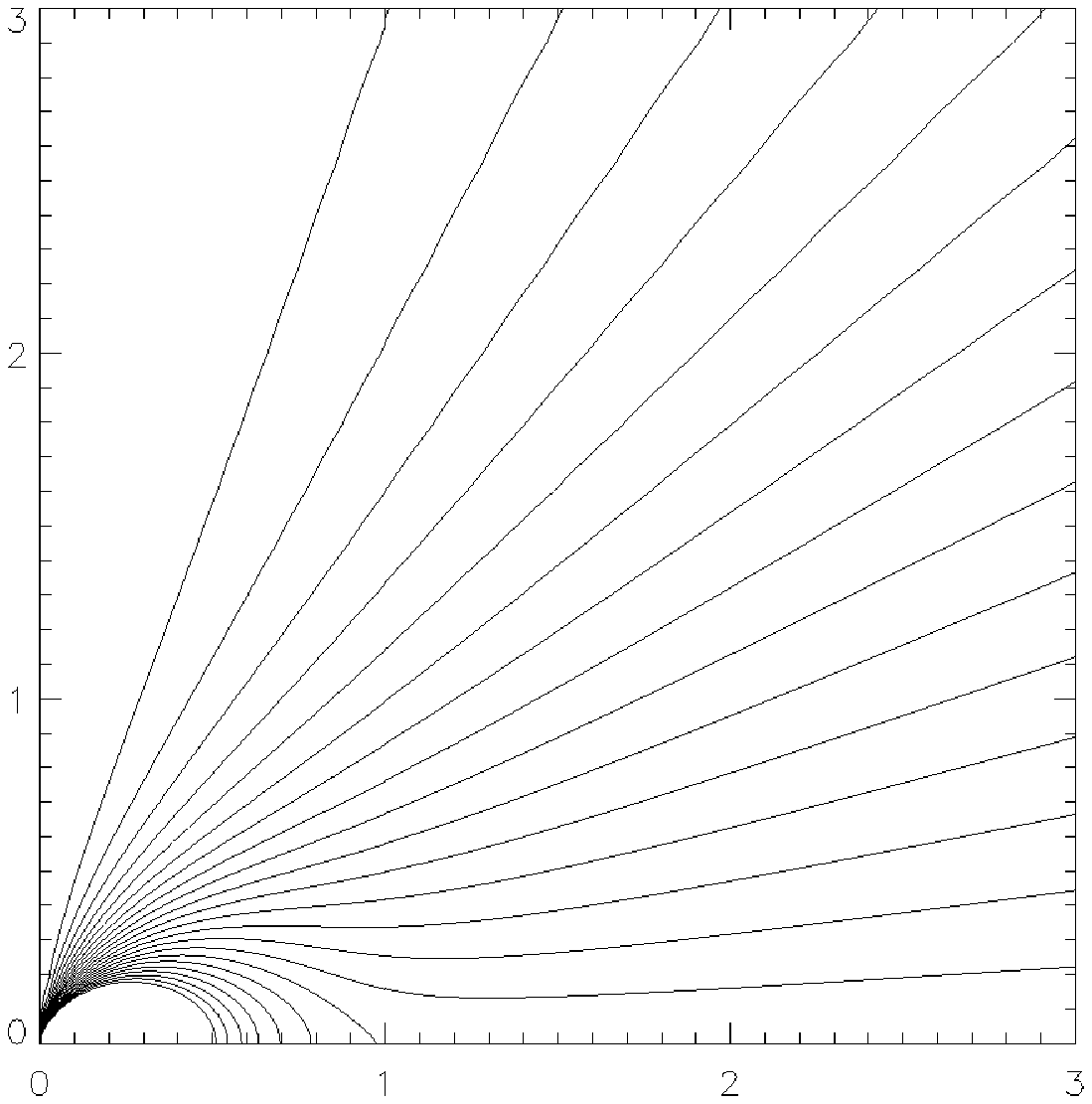}}}
\put(-10,270){\makebox(0,0){\tiny (b)}}
\end{picture}
\end{center}
\vskip -.15in
\caption{
Solutions to the pulsar equation
: a) CKF solution, b) unphysical but smooth solution without return current.}
\end{figure}


A number of pulsar magnetosphere models constructed as smooth solutions
at the light cylinder encountered regions with $E>B$ further away. Such regions
were interpreted as dissipation regions that made current closure via particles
crossing the flux surfaces (e.g., Mestel \& Shibata 1994). 
The CKF solution reports to have $E<B$ throughout, and, especially if verified,
serves as a counter-example where force-free electrodynamics
holds everywhere in the flow (potentially with the exception of the current sheet regions).
Although its existence is not a guarantee that it is stable and will be chosen by
a dynamical system, the CKF solution is very attractive on theoretical grounds, and 
is supported by observational absence of copious amounts of radiation that would 
emerge from a dissipation layer near $R_L$. 

\section{Time-dependent models}

The questions of stability and uniqueness of magnetospheric solutions are
better addressed with a time-dependent simulation than with iterative
steady-state searches. It also is tempting to simply
setup the appropriate dynamical equations with boundary conditions and let the
problem ``solve itself''. The limit of abundant supply of inertialess plasma
with large magnetization is well described by the force-free electrodynamics
equations (Blandford 2002):
\begin{eqnarray}
{1\over c} {\partial  {\bf E}\over \partial t}={\bf \nabla} \times {\bf B}&-&{4 \pi \over c} 
{\bf j}; \quad {1\over c} {\partial {\bf B}\over \partial t} =-{\bf \nabla} \times {\bf E}; \label{m1} \\
{\bf j}={c\over 4 \pi} ({\bf \nabla} \cdot {\bf E}) {{\bf E} \times {\bf B} \over B^2}&+&
{c\over 4 \pi} {({\bf B}\cdot {\bf \nabla}\times {\bf B}-{\bf E}\cdot {\bf \nabla}\times {\bf E}){\bf{B}}
\over B^2}.\label{m3}
\end{eqnarray}
The first two are just Maxwell's equations, while the third is a current prescription
that is derived from the force-free constraint and the ideal condition 
$\partial_t ({\bf E}\cdot {\bf B})=0$. The two terms
in (\ref{m3}) represent the transverse current due to advection of charge density 
perpendicular to field lines, and the longitudinal current respectively.
The system (\ref{m1}-\ref{m3}) is hyperbolic and can be evolved in time.
We solved these equations using finite-difference time-domain (FDTD)
method in spherical coordinates with rotating conducting sphere boundary conditions for 
aligned magnetic fields. Below I describe two test cases that illustrate the nature of 
force-free time-dependent electrodynamics: a spinup of a sphere with initially 
monopolar magnetic field, and a sphere with a dipole. The equations (\ref{m1}-\ref{m3})
support two wavemodes: fast (electromagnetic) and Alfven. In general, Alfven waves have
non-zero charge density (${\bf k}\cdot {\bf E}$), and they can transport charge along the 
magnetic field. Dynamics of these waves allows the magnetosphere to adjust to 
perturbations. For instance, in fig. 5a an Alfven wave is launched when a sphere with 
monopolar magnetic field starts rotating. Behind the wave front Goldreich-Julian charge
and current density are established, making the fieldlines corotate. The transient wave 
propagates out to infinity and simulation reproduces the analytic solution by 
Michel (1973a) and has no pathology at the light cylinder.
\vskip -.15in
\begin{figure}[hbt]
\unitlength = 0.0011\textwidth
\begin{center}
\hspace{1\unitlength}
\begin{picture}(200,200)(0,15)
\put(0,0){\makebox(200,200){ \epsfxsize=240\unitlength \epsfysize=240\unitlength
\epsffile{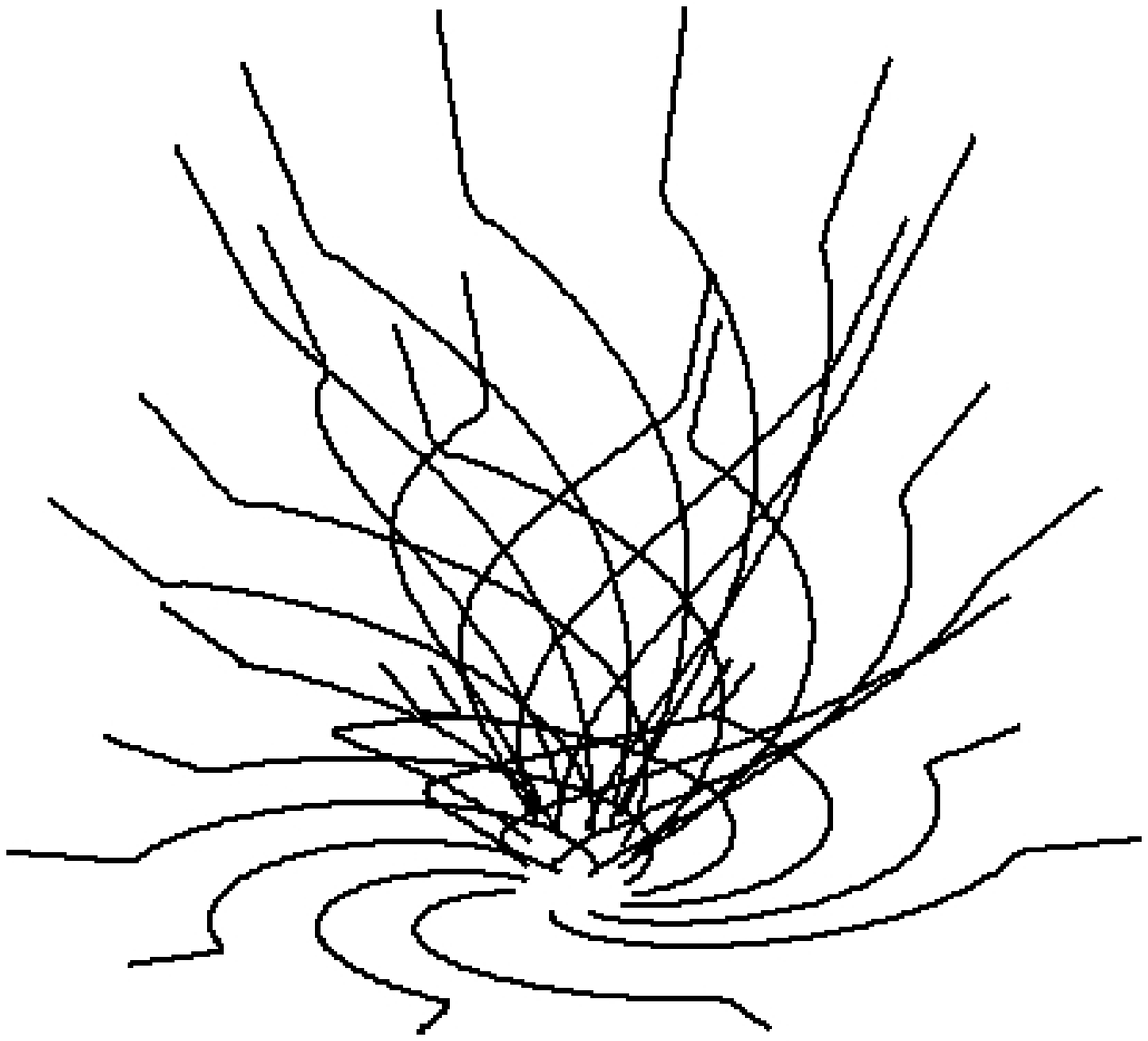}}}
\put(0,210){\makebox(0,0){\tiny (a)}}
\end{picture}
\hspace{1\unitlength}
\begin{picture}(140,200)(0,15)
\put(0,0){\makebox(140,200){\epsfxsize=210\unitlength \epsfysize=300\unitlength
\epsffile{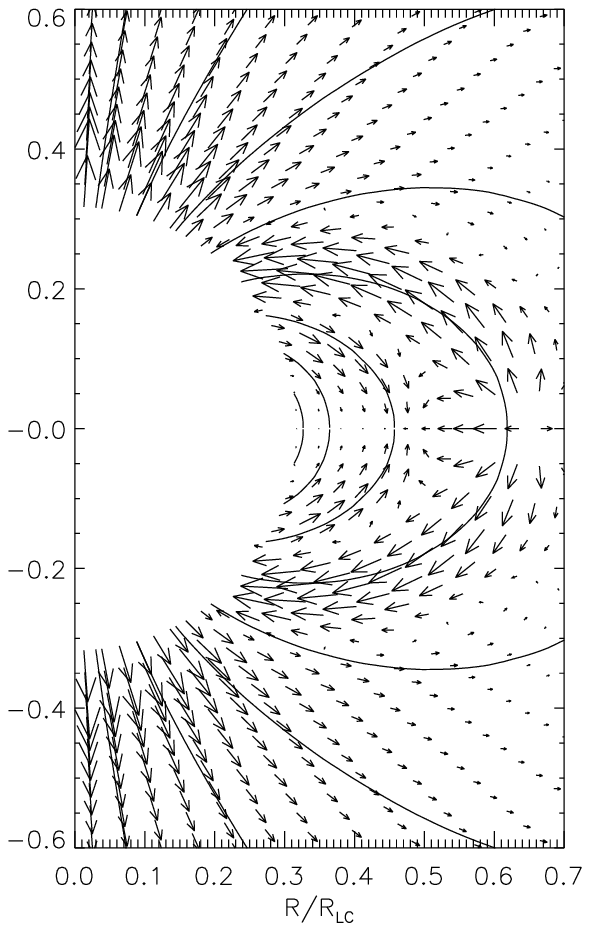}}}
\put(0,210){\makebox(0,0){\tiny (b)}}
\end{picture}
\hspace{30\unitlength}
\begin{picture}(140,200)(0,15)
\put(0,0){\makebox(140,200){\epsfxsize=210\unitlength \epsfysize=300\unitlength
\epsffile{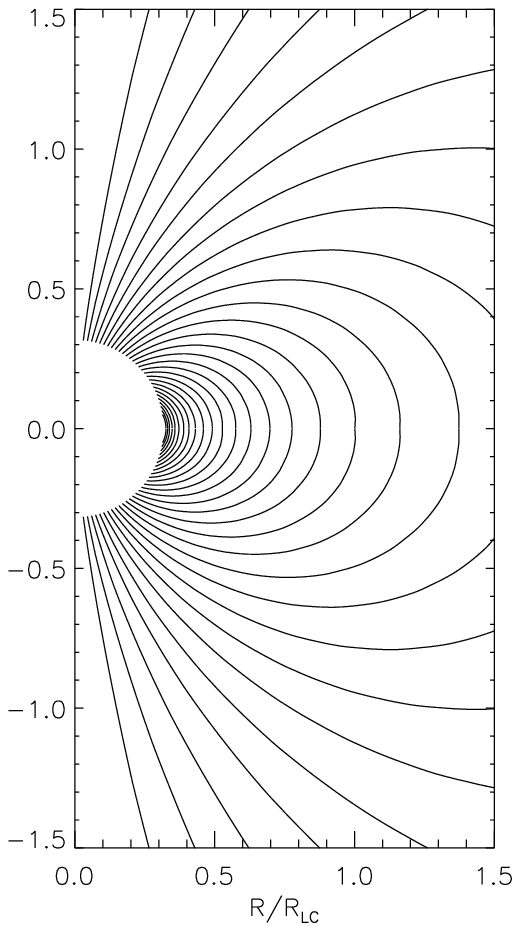}}}
\put(0,210){\makebox(0,0){\tiny (c)}}
\end{picture}
\hspace{1\unitlength}
\begin{picture}(140,200)(0,15)
\put(0,0){\makebox(140,200){\epsfxsize=210\unitlength \epsfysize=300\unitlength
\epsffile{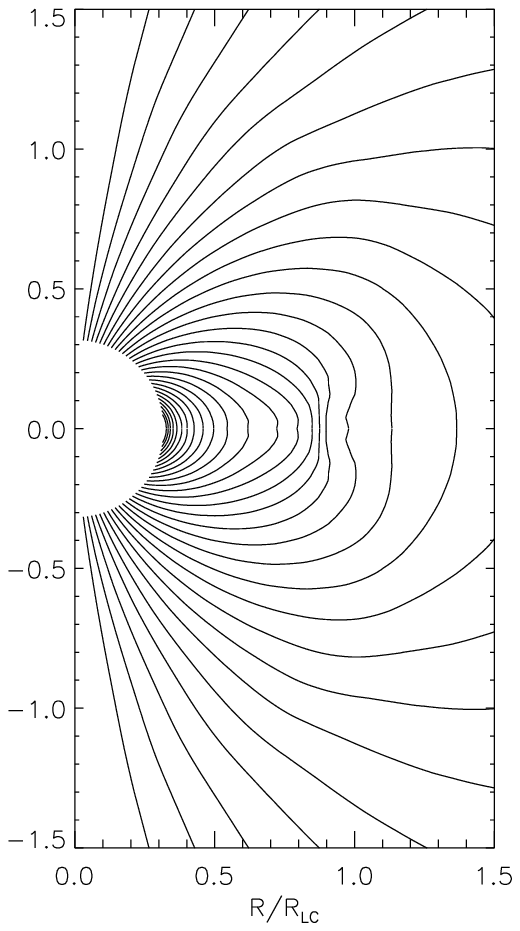}}}
\put(0,210){\makebox(0,0){\tiny (d)}}
\end{picture}
\hspace{1\unitlength}
\begin{picture}(140,200)(0,15)
\put(0,0){\makebox(140,200){\epsfxsize=210\unitlength \epsfysize=300\unitlength
\epsffile{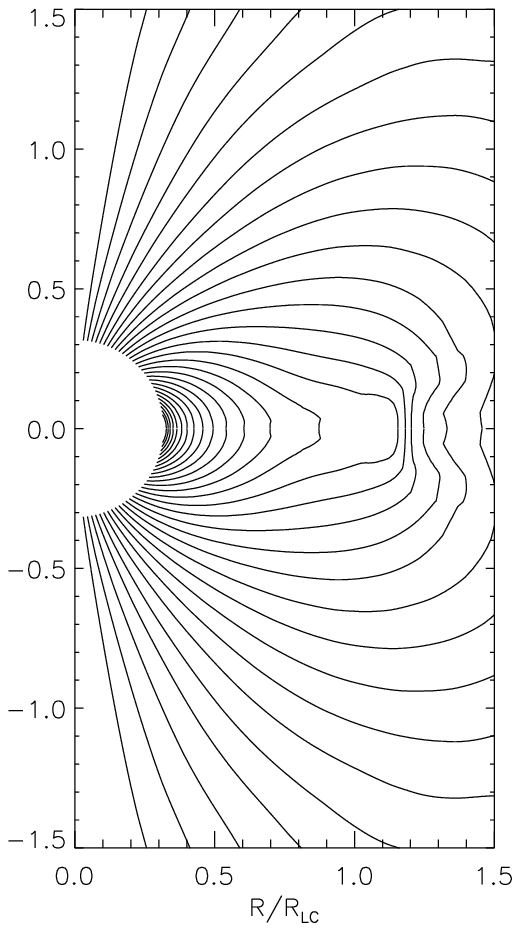}}}
\put(0,210){\makebox(0,0){\tiny (e)}}
\end{picture}
\end{center}
\caption{
a) Alfven wave from a rotating magnetic monopole; b) Snapshot of formation of 
return current for dipole field; c-e) Evolution of poloidal field for rotating dipole. }
\end{figure}

The situation is more involved for a rotating dipole. Initial evolution (shown in fig. 5b)
produces Alfven waves which carry outwards current on all field lines (for simplicity ${\bf \mu}$ and $\bf{\Omega}$ are counteraligned in this example). Because Alfven waves follow the 
field lines, poloidal currents from opposite hemispheres start to cancel each other in the 
closed zone. As signals from higher latitudes cancel, the closed zone expands outwards, now
polarized with the Goldreich-Julian charge density.
The currents do not cancel at the edge of the closed zone, and this is where a return
current sheath develops (fig. 5b). Corotation of the closed zone produces toroidal current, 
and the structure of the poloidal fieldlines, shown in figs. 5c-e, starts to change -- the field
lines are pulled out by the electromagnetic stress, and more flux is passing through the light
cylinder compared to the unperturbed dipole (as in CKF solution).  As the fieldlines are 
stretched horizontally, the net magnetic
field in the equator is decreasing. After $1/4$ of rotational period the total field 
in the equator at the light cylinder actually reaches zero, at which point the simulation 
breaks (note the denominator of (\ref{m3})). In the process of opening the field lines, the 
system tries to spontaneously form a 
current sheet in the equator separating the opposite toroidal and poloidal magnetic fields.
Such a current sheet is likely to be non-force-free, and requires the return of resistivity
and pressure to be supported. It is still an open question whether it is possible to modify
the set (\ref{m1}-\ref{m3}) by adding a prescription for resistivity or pressure effects to 
incorporate current
sheets into otherwise ideal force-free simulation. An interesting conceptual question is 
whether one can always smoothly pass 
from one ideal force-free configuration to another without invoking non-ideal effects.
A simple example is two CKF-type magnetospheres of different spin. Continuous transformation 
between them by changing the spin of the star involves converting some open flux to closed 
or vice-versa, and hence suggests the need for resistivity.
Thus, time-dependent evolution of force-free ideal equations suggests that extra physics
needs to be included to establish the steady state solution for the pulsar case
(note that this extra physics would likely act only in the current sheet and is different 
from an extended dissipation zone as in Mestel \& Shibata (1994)).

So, where do we go from here? In this brief review I outlined several directions of numerical
attack on the problem of global structure of pulsar magnetospheres. None have yielded a 
conclusive answer, although all signs for the aligned rotator with ample plasma supply 
are pointing to a CKF-type solution. I am optimistic that the CKF solution 
will eventually get confirmed, or a solution similar to CKF should exist. Especially promising
is the development of time-dependent global models of magnetosphere, as they have huge observational
implications. It is also the only way to address the problem of oblique rotators.
Once the right set of equations with enough physics is postulated, it is a matter 
of time and a lucky numerical procedure before it will get solved. One set of equations that
certainly has enough physics is relativistic MHD; however, RMHD becomes stiff and difficult
to solve numerically in the high magnetization (force-free) limit, so some sort of hybrid
scheme is probably required. 


\begin{references}
\reference Blandford, R. D. 2002, astro-ph/0202265
\reference Contopoulos, J., Kazanas, D. \& Fendt, C. 1999, \apj, 511, 351 (CKF)
\reference Goldreich, P. \& Julian, W. H. 1969, \apj, 160, 971
\reference Holloway, N. J., 1973, Nature Phys. Sci., 246, 6
\reference Krause-Polstorff, J. \& Michel, F. C. 1984, \mnras, 213, 43
\reference Krause-Polstorff, J. \& Michel, F. C. 1985, A\&A, 144, 72 
\reference Mestel, L. \& Shibata, S. 1994, \mnras, 271, 621
\reference Michel, F.~C. 1973a, \apj, 180, L133 and 1973b, \apj, 180, 207 
\reference Ogura, J. \& Kojima, Y. 2003, Progr. Theor. Phys., 109, 619
\reference Petri J., Heyvaerts J. \& Bonazzola, S. 2002a, \aap, 384, 414
\reference Petri J., Heyvaerts J. \& Bonazzola, S. 2002b, \aap, 387, 520
\reference Smith, I. A., Michel, F. C., Thacker, P. D. 2001, MNRAS, 322, 209 
\reference Spitkovsky, A. \& Arons, J. 2002, ASP Conf. Ser., 271, 81 (SA)
\end{references}
\end{document}